\documentclass[aps,prx,twocolumn,superscriptaddress,showpacs,floatfix]{revtex4-1}
\usepackage{amsmath,amssymb,graphicx}

\usepackage[utf8]{inputenc}
\usepackage[T1]{fontenc}
\usepackage{xcolor}

\IfFileExists{newtxtext.sty}
   {\usepackage{newtxtext,newtxmath}}
   {\IfFileExists{stix.sty}
      {\usepackage{stix}}
      {\IfFileExists{mathptmx.sty}
      {\usepackage{mathptmx}}{} } }

\usepackage{textcomp}

\usepackage{bm}

\IfFileExists{siunitx.sty}{\usepackage{booktabs,siunitx}}{}

\pdfoutput=1
\usepackage{color}
\definecolor{LinkColor}{rgb}{0.256,0.439,0.588}
\usepackage{hyperref}
\hypersetup{
   pdfauthor={Zi-Hong Liu, Xiao Yan Xu, Yang Qi, and Zi Yang Meng},
   pdftitle={Functional quantum Monte Carlo Method},
   colorlinks=true,
   citecolor=LinkColor,
   linkcolor=LinkColor,
   urlcolor=LinkColor
}

\renewcommand{\vec}[1]{\mathbf{#1}}

\usepackage{pifont}

\begin{document}

\title{EMUS-QMC: Elective Momentum Ultra-Size Quantum Monte Carlo Method}

\author{Zi Hong Liu}
\thanks{These authors have contributed equally to this work.}
\affiliation{Beijing National Laboratory for Condensed Matter Physics and Institute
of Physics, Chinese Academy of Sciences, Beijing 100190, China}
\affiliation{School of Physical Sciences, University of Chinese Academy of Sciences, Beijing 100190, China}
\author{Xiao Yan Xu}
\thanks{These authors have contributed equally to this work.}
\affiliation{Department of Physics, Hong Kong University of Science and Technology, Clear Water Bay, Hong Kong, China}

\author{Yang Qi}
\email{qiyang@fudan.edu.cn}
\affiliation{Center for Field Theory and Particle Physics, Department of Physics, Fudan University, Shanghai 200433, China}
\affiliation{State Key Laboratory of Surface Physics, Fudan University, Shanghai 200433, China}
\affiliation{Collaborative Innovation Center of Advanced Microstructures, Nanjing 210093, China}

\author{Kai Sun}
\affiliation{Department of Physics, University of Michigan, Ann Arbor, MI 48109, USA}

\author{Zi Yang Meng}
\email{zymeng@iphy.ac.cn}
\affiliation{Beijing National Laboratory for Condensed Matter Physics and Institute of Physics, Chinese Academy of Sciences, Beijing 100190, China}
\affiliation{CAS Center of Excellence in Topological Quantum Computation and School of Physical Sciences, University of Chinese Academy of Sciences, Beijing 100190, China}
\affiliation{Songshan Lake Materials Laboratory, Dongguan, Guangdong 523808, China}

\begin{abstract}
One bottleneck of quantum Monte Carlo (QMC) simulation of strongly correlated electron systems lies at the scaling relation of computational complexity with respect to the system sizes. For generic lattice models of interacting fermions, the best methodology at hand still scales with $\beta N^3$ where $\beta$ is the inverse temperature and $N$ is the system size. Such scaling behavior has greatly hampered the accessibility of the universal infrared (IR) physics of many interesting correlated electron models at (2+1)D, let alone (3+1)D. To reduce the computational complexity, we develop a new QMC method with inhomogeneous momentum-space mesh, dubbed elective momentum ultra-size quantum Monte Carlo (EQMC) method. Instead of treating all fermionic excitations on an equal footing as in conventional QMC methods, by converting the fermion determinant into the momentum space, our method focuses on fermion modes that are directly associated with low-energy (IR) physics in the vicinity of the so-called hot-spots, while other fermion modes irrelevant for universal properties are ignored. As shown in the manuscript, for any cutoff-independent quantities, e.g. scaling exponents, this method can achieve the same level of accuracy with orders of magnitude increase in computational efficiency. We demonstrate this method with a model of antiferromagnetic itinerant quantum critical point, realized via coupling itinerant fermions with a frustrated transverse-field Ising model on a triangle lattice. The system size of $48 \times 48 \times 32$ ($L\times L\times\beta$, almost 3 times of previous investigations) are comfortably accessed with EQMC.
With much larger system sizes, the scaling exponents are unveiled with unprecedentedly high accuracy, and this result sheds new light on the open debate about the nature and the universality class of itinerant quantum critical points.
\end{abstract}

\date{\today}

\maketitle

\section{Introduction}
\label{sec:intro}

As an unbiased numerical method, determinantal quantum Monte Carlo (DQMC) is widely used to study sign-problem-free interacting-fermion systems~\cite{BSS1981,Hirsch_1983,White1989,Assaad1996,Assaad1999,AssaadEvertz2008,Meng2010,He2016a,He2016b}. Despite its great success, for systems with large correlation lengths, such as quantum critical systems, it is still highly challenging for this method to accurately reveal the fate of the system at the thermodynamic limit, because of the sharp rise in computational complexity for systems with large sizes, which scales as $O\left(\beta N^3\right)$ with $\beta$ being the inverse temperature and $N$ being the volume of the lattice~\cite{White1989,AssaadEvertz2008}.
Although polynomial, this complexity becomes prohibitive for accessing low-temperature physics in large systems, which are nevertheless necessary for the endeavor such as itinerant quantum criticality in correlated electron systems~\cite{Hertz1976,Millis1993,Moriya1985,Loehneysen2007}.

To address this challenge, we develop a new QMC method dubbed elective momentum ultra-size quantum Monte Carlo (EMUS-QMC, or EQMC in short), which applies generically to any itinerant quantum critical systems, although it is optimized for systems with finite wave-length instabilities (e.g. charge density waves or spin density waves). This method drastically reduces the computational costs and thus can be used to simulate larger systems in order to reveal long-distance physics and universal properties near a quantum critical point. The key to this method lies in the concept of renormalization group (RG) and the following observation. In DQMC, computational costs mainly come from handling the fermionic modes. However, in an itinerant fermionic systems, fermionic excitations associated with low-energy excitations, which dominate the universal behaviors at infrared (IR), only inhabit a small part of the Brillouin zone (BZ). All other fermionic modes, which are associated with high-energy physics, are only needed for ultraviolet (UV) completion and are irrelevant as far as universal quantum-critical phenomena are concerned. In conventional DQMC simulations, all fermionic modes are treated on an equal footing, and thus for universal properties at IR, e.g.  scaling laws, a large portion of the computational resource is ``wasted'' on the high-energy UV modes, irrelevant in the sense of RG~\footnote{Similar understanding has been realized in the recent work by Lang and Laeuchli~\cite{Lang2018}, where by employing long-range hopping, the fermionic modes at IR can be extended to most of the BZ, thereby minimizes the finite size effect at the quantum criticality of DQMC simulations.}.

In the EMUS-QMC method that we developed, the fermion determinant is computed in the momentum space. Instead of treating all fermionic modes equivalently, we select momentum points within certain patches of the BZ, in which fermionic excitations directly contributes to low-energy (IR) physics. When we compute the fermion determinant, only fermionic modes with momenta inside these patches are taken into account, while all other fermionic degrees of freedom are ignored. As mentioned above, these ignored fermion modes are irrelevant for universal behaviors at IR, and thus our method will produce the same value for any physics observables independent of UV cut-offs, e.g., scaling exponents. Because this procedure reduces the number of fermionic degrees of freedom, the computational complexity can be greatly reduced from $O\left(\beta N^3\right)$ down to $O\left(\beta N_f^3\right)$, where $N$ is the volume of the whole BZ and $N_f$ is the volume of the momentum patch that are kept in the simulation (measured in the number of fermion modes). Speedup, to the level of $10^{3}$, can be  easily achieved in this method as it is easy to reach $\frac{N}{N_f}\sim 10$. In fact, for the model studied in Sec.~\ref{sec:demo}, we showed that reliable and accurate results can be obtained even for $\frac{N}{N_f}=36$, which implies an even more drastic reduction in computational cost.

Same as many other RG-based techniques, our method only focuses on cut-off independent universal properties. For quantities that are sensitive to UV physics, e.g. the value of the critical temperature or critical coupling strength, our technique is not expected to generate the same value  in comparison with the full-momentum-space DQMC simulations. Examples of this type are also shown below.

Our EQMC method is particularly suitable for studying quantum criticality in correlated itinerant electron systems, which is a subject with great theoretical and experimental significance~\cite{Hertz1976,Millis1993,Moriya1985,Stewart2001,Chubukov2004,Belitz2005,Loehneysen2007,Chubukov2009}, and plays a vital role in the understanding of anomalous transport, strange metal and non-fermi-liquid behaviors~\cite{Metzner2003,Senthil2008,Holder2015,Metlitski2015,Xu2017} in heavy-fermion materials~\cite{Custers2003,Steppke2013}, Cu- and Fe-based high-temperature superconductors~\cite{ZhangWenLiang2016,LiuZhaoYu2016,Gu2017}, as well as the recently discovered pressure-driven quantum critical
point (QCP) between magnetic order and superconductivity in transition-metal monopnictides, CrAs~\cite{Wu2014}, MnP~\cite{Cheng2015}, CrAs$_{1-x}$P$_x$~\cite{Matsuda2018} and other Cr/Mn-3d electron systems~\cite{Cheng2017}. Despite the extensive theoretical efforts in the past decades~\cite{Hertz1976,Millis1993,Moriya1985,Abanov2003,Abanov2004,Metlitski2010b,Sur2016,Schlief2017}, the problem of itinerant quantum criticality is still open and among the most difficult ones in strongly correlated electron systems, due to its nonperturbative nature.

Admittedly, the recent development in sign-problem-free DQMC methods paves the way to sharpen our understanding on this open question. Coupling a Fermi surface (FS) with various bosonic critical flucutations, many itinerant QCPs, including Ising-nematic~\cite{Schattner2015a,Lederer2016}, ferromagnetic~\cite{Xu2017}, charge density wave~\cite{ZXLi2015,ChuangChen2018a}, spin density wave~\cite{Berg12,ZXLi2016,Schattner2015b,Gerlach2017,ZHLiu2017,ZiHongLiu201808} and interaction-driven topological phase transitions~\cite{Xu2016a,Assaad2016,He2017} have been studied.
However, the true scaling behaviors of the itinerant QCPs are yet to be explored, because the aforementioned $\beta N^3$ complexity prohibits the simulations to reach the true thermodynamic limit, even if with help of the very recent advances such as self-learning Monte Carlo method~\cite{liu2016self,liu2016fermion,Xu2016self,Nagai2017,ChuangChen2018a}.
In this regard, the EQMC method developed in this work, provides a systematic way to reach larger system sizes, and consequently allowing us to access the genuine scaling behaviors in the IR limit for itinerant quantum criticality.

The rest of the paper is organized as follows: the general formulation of EQMC method is laid out in Sec.~\ref{sec:method}, and in Sec.~\ref{sec:demo}, it is applied to a previously studied model~\cite{ZHLiu2017} for benchmarking purpose, and with much larger system sizes simulated with EQMC, our previous conclusions are more firmly established. Finally, remarks on future applications of EQMC are provided in Sec.~\ref{sec:conclusion}.

\section{EMUS quantum Monte Carlo Method}
\label{sec:method}

\subsection{General Ideas}
\label{sec:ideas}

In general, DQMC simulations study the following fermion-boson models,
\begin{equation}
\label{eq:Hgeneral}
H = H_f + H_{fb} + H_b,
\end{equation}
where $H_f$ contains a quadratic free-fermion model,
\begin{equation}
\label{eq:Hf}
H_f=-\sum_{ija}t_{ij}\left(c_{ia}^\dagger c_{ja}+\text{H.c.}\right);
\end{equation}
$H_{fb}$ describes the coupling between the fermions and bosonic modes, which is local and also quadratic in terms of fermion operator $c$ and $c^{\dagger}$s:
\begin{equation}
\label{eq:Hfb}
H_{fb}=\lambda\sum_ic_{ia}^\dagger M_{ab}c_{ib}\phi_i;
\end{equation}
$H_b$ describes a Hamiltonian of the bosonic modes $\phi_i$, which can take any generic form. Here, $a,b$ denotes a combination of quantum numbers, including orbits and spins, that labels different fermion species. The summation over repeated indices of fermion species are taken implicitly if not specified explicitly.
Such a model can either be obtained from an interacting fermion model, through introducing auxillary fields using a Hubbard-Stratonovish transformation~\cite{BSS1981,Hirsch_1983,AssaadEvertz2008}, or be constructed directly as a ``designer'' Hamiltonian~\cite{Xu2017,ZHLiu2017}.

\begin{figure}[htp!]
\centering
\includegraphics[width=\columnwidth]{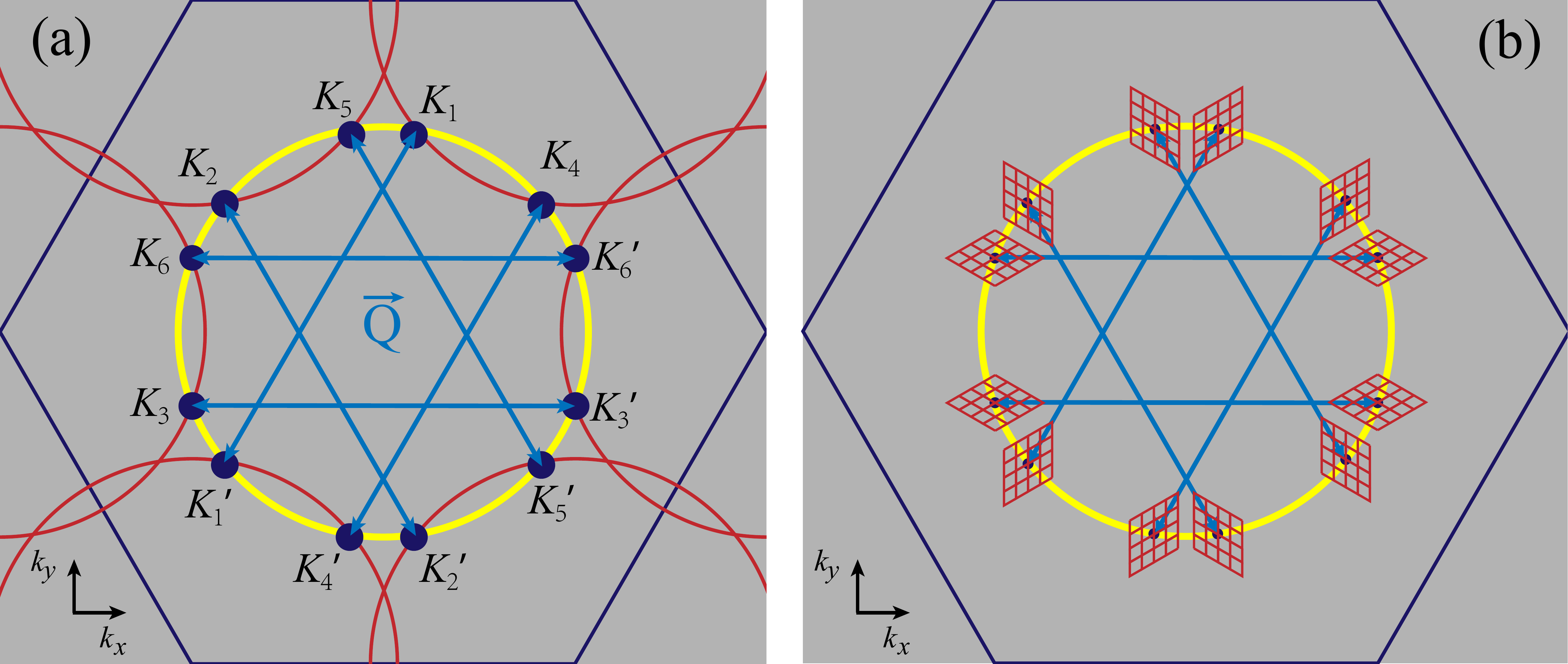}
\caption{(a) The bare FS of the $H_{f}$, here as in Sec.~\ref{sec:model}, the FS (yellow circle) is for a triangle lattice model. The folded FS (red circles), coming from translating the bare FS by momentum $\vec{Q}$ (blue arrows), which is the wave vector for the antiferromagnetic fluctuations. The folded FS contains Fermi pockets and hot spots (black dots). Two hot spots connected by momentum $\vec{Q}$ forms a hot spot pair and we label it by $\{\mathbf{K}_l,\mathbf{K'}_l\}$, with $l=1,2,\cdots,6$ in case of a triangle lattice model. (b) shows the inhomogeneous $\mathbf{k}$ mesh (red rhombus mesh) build around hot spots, the number of momentum points inside each mesh is denoted as $N_f$.}
\label{fig:FS}
\end{figure}

In DQMC simulations, configurations of the bosonic modes $\phi_i$ are stochastically sampled, with weights obtained through integrating out the fermion modes. Traditionally, real-space fermion modes, $c_{ia}$, are chosen as eigenstates for this computation. In EQMC, we instead treat fermion modes in the momentum space. To this end, we rewrite Eqs.~\eqref{eq:Hf} and \eqref{eq:Hfb} in momentum space,
\begin{equation}
\label{eq:Hfk}
H_f=\sum_k[\epsilon(k)-\mu]c_{ka}^\dagger c_{ka};
\end{equation}
and
\begin{equation}
\label{eq:Hfbk}
H_{fb}=\lambda\sum_{kk^\prime}c_{ka}^\dagger M_{ab}c_{k^\prime b}
\phi_{k-k^\prime}.
\end{equation}
Here, $\phi_k$ denotes the $k$-component of the Fourier transform of the bosonic field $\phi_i$:
\begin{equation}
  \label{eq:phik}
  \phi_k = \frac1N\sum_i\phi_i e^{-i\bm k\cdot\bm r_i}.
\end{equation}

Rewriting the same problem in the momentum space gives us the freedom of choosing arbitrary $k$ points in the momentum summation in Eqs.~\eqref{eq:Hfk} and \eqref{eq:Hfbk}.
When studying the low-energy and long-range physics, we choose IR fermion modes that are particularly relevant for this physics, and throw away other fermion modes without worrying a proper UV completion of a lattice model.
In particular, for studying fermionic QCPs, only fermion modes near the so-called ``hot spots'', where two patches of FSs are connected by the ordering wave vector $\mathbf{Q}$ [see Fig.~\ref{fig:FS}(a)], are relevant to the universality class of the QCPs~\cite{Metlitski2010b,Abanov2004}.
Therefore, we keep modes in patches around these hot spots, and neglect other modes in the BZ, as shown in Fig.~\ref{fig:FS}(b).
In this way, the number of fermion modes used in computing the effective weights is greatly reduced from the total size (or total volume) $N$ to the patch size (or patch volume) $N_f$, thus drastically leave the computational burden, while retaining the same IR physics of what a lattice DQMC simulation of system size $N$ can achieve.

When working with fermionic QCPs, the ratio $N_f/N$ can be chosen with the following principle. Naively, in a lattice model, the lattice constant $a$ provides a natural UV cutoff scale $\Lambda\sim\frac{2\pi}a$, which is also the size of the BZ.
However, in practice, in a lattice model with the dispersions shown in Fig.~\ref{fig:FS}(a), the cutoff $\Lambda$ can be reduced. Due to the complicated shape of the Fermi surface, the universal scaling behavior is achieved assuming a linearized Fermi surface around the hot spots.
Therefore, the size of the approximated linear portion of the Fermi surface provides the real UV cutoff $\Lambda$, which is much smaller than the size of the BZ.
$\Lambda$ also defines an ``optimal'' patch around each hot spot: when the patch size is larger than $\Lambda$, the cutoff scale is still $\Lambda$ instead of the patch size. Therefore, the computational cost is reduced, while retaining the same universal physics.
In other words, fermion modes outside the ``optimal'' patch do not contribute to the universal IR physics, and can be safely removed.
However, reducing the patch size beyond $\Lambda$ does not offer more acceleration: although the computational cost is reduced, the cutoff is also reduced proportionally, since the cutoff is now given by the patch size. Hence, to reach the same ratio between UV/IR scales, which is needed for studying universal scaling behaviors, a finer mesh of momentum points in the patch must be used, which contains the same amount of fermion modes and thus requires no less amount of computational time. Therefore, the ratio $N/N_f$ should be chosen such that the patch has the ``optimal'' size.

We emphasize that by removing fermion modes outside of IR patches, the nonuniversal behavior of the model is quantitatively varied.
In other words, the EQMC method is simulating a \emph{different} model, which shares the same universal IR physics, including the scaling behaviors at the QCP, as the original lattice model.
This is clearly demonstrated in the example studied in Sec.~\ref{sec:demo}.

In EQMC, working in the momentum space means that we can no longer take advantage of the locality in Eq.~\eqref{eq:Hfb}, and therefore cannot efficiently perform local updates in the conventional DQMC~\cite{AssaadEvertz2008}.
In lattice-based DQMC, the local update~\cite{BSS1981,Hirsch_1983,Hirsch_1985,AssaadEvertz2008}, which tries to flip the bosonic spins $s_{i,\tau}$ one by one through the space-time lattice of volume $\beta N$, can benefit from the so-called ``fast update''. The acceptance ratio of such a flip involves a ratio of two determinants before and after the flip.
The local nature of the update enables one to perform a fast update with a complexity $O(1)$ to compute this ratio, and a complexity $O(N^2)$ to update the Green's function if the flip is accepted~\cite{BSS1981}, although generally
the computational complexity for evaluating a determinant is $O(N^3)$.
[Since filpping each spin costs $O(N^2)$, a full update that flips spins on the order of space time volume $\beta N$ costs $O(\beta N^3)$, which is the well-known complexity of DQMC.]
In EQMC, due to the dense nature of the coupling in Eq.~\eqref{eq:Hfbk}, one can no longer use the fast update algorithm, and a local update would cost $\beta N\cdot O(\beta N_f^3)$.
Fortunately, there are global-update algorithms one can choose, including the cumulative update in self-learning Monte Carlo method developed recently by some of us~\cite{liu2016fermion,Xu2016self}. This gives rise to the complexity $O(\beta N_f^3)$ for computing the fermion determinant in a full update, which dominants the total computational cost of EQMC (see the detailed discussion in Sec.~\ref{sec:complexity}). Since $N_f$ can be much smaller than $N$, speedup of $(\frac{N}{N_f})^3 \sim 10^3$ with $\frac{N}{N_f} \sim 10$ can be easily foreseen.



\subsection{Steps of the Algorithm}
\label{sec:steps}

We now describe the details of the EQMC algorithm. Like all DQMC simulations, it samples through bosonic-field configurations $\{\phi_{i,\tau}\}$, with the following weights,
\begin{equation}
  \label{eq:Wphi}
  W[\phi] = W_b[\phi]\det(\mathbf I + \mathbf B(\beta,0;\phi)).
\end{equation}
The weight has two parts: $W_b[\phi]$ denotes a bosonic weight determined by the Hamiltonian $H_b$ in Eq.~\eqref{eq:Hgeneral}, and $\det(\mathbf I + \mathbf B(\beta,0;\phi))$ is a fermion determinant obtained by integrating out the fermion modes in the patches.
Here, $\mathbf I$ denotes the identity matrix; and $\mathbf B(\beta, 0;\phi)$ is a short form for the matrix product of $\mathbf{B}^M \mathbf{B}^{M-1} \cdots \mathbf{B}^1$,  where the matrix at time slice $\tau$ is $\mathbf{B}^\tau = \exp({\Delta \tau \mathbf{K}})\exp({\mathbf{V}[\phi}])$, with $\mathbf{K}$ the kinetic-energy matrix of the bare system in $H_f$ in Eq.~\eqref{eq:Hfk}:
$K_{ka,k^\prime b} = [\epsilon(\bm k)-\mu]\delta_{kk^\prime}\delta_{ab}$, and $\mathbf{V}[\phi]$ the fermion-boson coupling in $H_{fb}$ in Eq.~\eqref{eq:Hfbk}: $V_{ka,k^\prime b}=\lambda M_{ab}\phi_{k-k^\prime}$.

The bosonic weight $W_b[\phi]$ can be computed in standard ways, as in bosonic QMC simulations~\cite{YCWang2017}.
Computing the fermionic determinant, on the other hand, takes three steps:
(i) at each time slice, the configuration $\phi_{i,\tau}$ is Fourier transformed into momentum-space components $\phi_{k,\tau}$.
(ii) the interaction matrix $\mathbf{V}[\phi]$ is constructed using $\phi_{k,\tau}$.
(iii) the determinant $\det(\mathbf I + \mathbf B(\beta,0;\phi))$ is computed using the matrix $\mathbf{K}$, with all its elements fixed, and the matrix $\mathbf{V}[\phi]$, which is varying at each time slice.

The fermion matrices $\mathbf{K}$ and $\mathbf{V}[\phi]$ are generally $N_f\times N_f$ matrices.
For the problem of fermion QCP with multiple pairs of hot-spot patches, an approximation can be made such that the matrices have a block-diagonalized structure, which further reduces the computational cost.
In such a problem illustrated in Fig.~\ref{fig:FS}, there are 12 hot spots, and they are divided into 6 pairs, each connected by the ordering wave vector $\mathbf{Q}$.
Taking two momenta $\mathbf{k}$ and $\mathbf{k^\prime}$ from the patches around one pair of hot spots, the momentum transfer $\mathbf{k^\prime}-\mathbf{k}\simeq \mathbf{Q}$ corresponds to a low-energy bosonic mode.
On the other hand, the momentum transfer $\mathbf{k^\prime}-\mathbf{k}$ between two other momenta, which is not close to $\mathbf{Q}$, corresponds to a high-energy bosonic mode.
Therefore, we make an approximation to only include in $\mathbf{V}$ the scattering within a pair of hot-spot patches connected by $\mathbf{Q}$, and ignore other scattering processes. Under this approximation, the matrix $\mathbf{V}$ becomes block-diagonalized, with $2N_f\times 2N_f$ blocks, where $2N_f$ is the total number of modes in the pair of patches. Since the matrix $\mathbf{K}$ is already diagonal in momentum space, the resulting matrices $\mathbf B^\tau$ is also block-diagonalized.
This block-diagonalized structure allows the determinant of different blocks to be computed individually, thus saving both computational time and memory.

\subsection{Complexity of the Algorithm}
\label{sec:complexity}

The computational complexity of the EQMC algorithm can be determined following the steps outlined in Sec.~\ref{sec:steps}.
We first consider the complexity of computing the weight $W[\phi]$ in Eq.~\eqref{eq:Wphi} for a given configuration $\{\phi_i\}$.
The complexity of computing the bosonic weight $W_b[\phi]$ depends on the form of $H_b$.
For a local Hamiltonian with finite-ranged interaction, this can be done with complexity $O(\beta N)$.
Next, the Fourier transform from $\phi_i$ to $\phi_k$ is done using the fast Fourier transform (FFT) algorithm, which costs $O(N\log N)$ for each time slice, and $O(\beta N\log N)$ in total.
Comparing to this, the cost of computing $W_b[\phi]$ is much smaller and thus can be neglected.
Then, constructing the matrices $V$ costs $O(N_f^2)$ for each time slice, and $O(\beta N_f^2)$ in total.
However, this is negligible comparing to the cost of computing the determinant $\det[\mathbf{I} + \mathbf{B}(0,\beta;\phi)]$, which costs $O(\beta N_f^3)$.
Therefore, the total computational cost for computing the weight of a given bosonic-field configuration $\{\phi_i\}$ is $O(\beta N\log N) + O(\beta N_f^3)$.

In typical models, the ``optimal'' patch occupies about 1\%-20\% of the total BZ. Therefore, in practice, we usually have $N/N_f<100$.
With this parameter range, the second term $O(\beta N_f^3)$ is much larger than the first term $O(\beta N\log N)$ for a large system.
Therefore, we can safely ignore the first term, and use $O(\beta N_f^3)$ as the complexity of computing the weight $W[\phi]$.


The complexity of generating an uncorrelated configuration in EQMC is also $O(\beta N_f^3)$,
assuming that we have an effective global-update algorithm satisfying the following assumptions:
(i) the computational cost of generating a new configuration $\{\phi_i\}$ is small comparing to $O(\beta N_f^3)$;
(ii) the autocorrelation time, measuring the number of steps in the generated Markov chain between two uncorrelated configurations, is of order one.
Examples of such update algorithms, in the form of the cumulative update scheme in the recently developed self-learning Monte Carlo method~\cite{liu2016fermion,Xu2016self}, will be given in Sec.~\ref{sec:results}, in the context of a concrete model.
Comparing to the cost of lattice-based DQMC, $O(\beta N^3)$, the EQMC thus offers an  very promising acceleration factor of $(\frac{N}{N_f})^3$.

\section{Demonstration of EQMC}
\label{sec:demo}

\subsection{Model}
\label{sec:model}

\begin{figure*}[htp!]
\centering
\includegraphics[width=\textwidth]{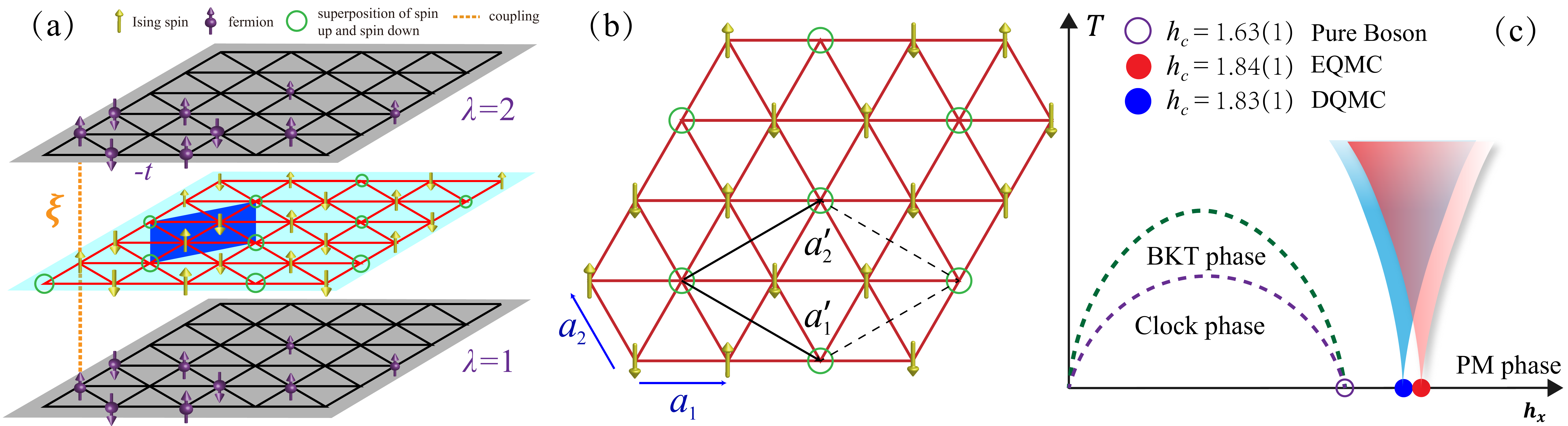}
\caption{(a) Illustration of the triangle lattice model in Sec.~\ref{sec:model}. Fermions reside on two of the layers ($\lambda$ = 1,2) with intra-layer nearest-neighbor hopping $t$. The middle layer is composed of Ising spins $s^{z}_{i}$, subject to nearest-neighbor antiferromagnetic Ising coupling $J$ and a transverse magnetic field $h$. Between the layers, an onsite Ising coupling is introduced between fermion and Ising spins ($\xi$). The shaded area in the middle layer stands for the enlarged unit cell of the clock phase before the Ising spins are polarized to the $s^{x}$ direction. (b) Illustration of the spin arrangement of the clock phase when $h<h_c$. The $\mathbf{a}_1$, $\mathbf{a}_2$ are the lattice vector of the original triangle lattice and $\mathbf{a'}_1$, $\mathbf{a'}_2$ are the lattice vector of the clock phase with enlarged unit cell. (c) Semi-quantatitive phase diagram. The dashed lines mark the phase boundaries of the pure bosonic model $H_{b}$, with a QCP (open magenta dot) at $h_c=1.63(1)$~\cite{Isakov2003,YCWang2017}. After coupling with fermions, the QCP shifts to higher values. The blue solid dot is the QCP of the originial model in Ref.~\onlinecite{ZHLiu2017} with homogeneous grid ($h_c=1.83(1)$), and the blue area is the quantum critical region where Hertz-Mills-Moriya scaling has been observed. The red solid dot is the QCP of this study ($h_c=1.84(1)$), The position of the QCP obtained from DQMC and EQMC is very close, and also the scaling behavior inside the quantum critical region (the red shaded area) is consistent with that observed in the blue shaded area. The EQMC scheme can comfortably capture the IR physics of such antiferromagnetic itinerant QCP, with much larger system sizes, $48\times 48$, compared with previous study~\cite{ZHLiu2017} $30\times30$ where great computational efforts have been spent.}
\label{fig:model}
\end{figure*}

In this section, we use the model studied in Ref.~\cite{ZHLiu2017} to benchmark and demonstrate the power of the EQMC method. The Hamiltonian of the model can be organized into the form of Eq.~\eqref{eq:Hgeneral}, with each part given by
\begin{align}
  \label{eq:HfM}
H_{f}&=-t\sum_{<ij>,\lambda,\sigma}(c^{\dagger}_{i,\lambda,\sigma}c_{j,\lambda,\sigma} +h.c.)-\mu\sum_{i}n_{i}, \\
\label{eq:HbM}
H_{b}&=J\sum_{<ij>}s^{z}_{i}s^{z}_{j}-h\sum_{i}s^{x}_{i}, \\
H_{fb}&=-\xi\sum_{i}s^{z}_{i}(\sigma^{z}_{i,1}+\sigma^{z}_{i,2}).
\end{align}
As shown in Fig.~\ref{fig:model} (a), fermions, subject to intra-layer nearest-neighbor hopping $t$ and chemical potential $\mu$, reside on two of the layers $\lambda=1,2$. The middle layer is composed of Ising spins $s^{z}_{i}$ with frustrated antiferromagnetic Ising coupling $J>0$ and a transverse magnetic field $h$ along $s^{x}$. Fermions and Ising spins are coupled together via an inter-layer onsite Ising coupling $H_{fb}$, where $\sigma^{z}_{i,\lambda}=\frac{1}{2}(c^{\dagger}_{i,\lambda,\uparrow}c_{i,\lambda,\uparrow}-c^{\dagger}_{i,\lambda,\downarrow}c_{i,\lambda,\downarrow})$ is the fermion spin along $z$. We set $t=1$, $J=1$, $\mu=-0.5$ (electron density $\langle n_{i,\lambda}\rangle \sim 0.8$) and leave $h$ and $\xi$ as control parameters.

It is well-known that $H_{b}$, describing a frustrated triangular-lattice transverse-field Ising model, has extensive ground state degeneracy at $h=0$. At finite $h$, this degeneracy is lifted by the quantum order-by-disorder effect, resulting in an ordered ground state with a clock pattern~\cite{Moessner2001}, as shown in Fig.~\ref{fig:model} (b). The clock phase spontaneously breaks the translational symmetry~\cite{Isakov2003,YCWang2017} and thus has an enlarged unit cell. This phase is characterized by a complex order parameter $me^{i\theta}=m_{1}+m_{2}e^{i4\pi/3}+m_{3}e^{-i4\pi/3}$ where $m_{\alpha}=\frac{3}{N}\sum^{N/3}_{i=1} s^{z}_{i,\alpha}$ with $\alpha=1,2,3$ representing magnetization of the three sublattices of the $\sqrt{3}\times\sqrt{3}$ enlarged superlattice. In the momentum space, this order parameter has a finite wave-vector $\mathbf{Q}=(\frac{2\pi}{3},\frac{2\pi}{\sqrt{3}})$, i.e., the corner of the hexagonal Brillouin zone as shown in Fig.~\ref{fig:FS}(a). Upon introducing quantum/classical fluctuations via increasing $h$ or $T$, the ordered phase can melt. The quantum melting is through a second-order quantum phase transition at $h_c=1.63(1)$ with an emergent U(1) symmetry~\cite{YCWang2017}. Because of this emergent continuous symmetry, despite that $H_{b}$ describes an Ising model, this quantum critical point belongs to the $(2+1)$D XY universality class and the thermal melting of the clock phase involves an intermediate BKT phase~\cite{YCWang2017}.

In the presence of the fermion-spin coupling, which is relevant in the RG sense, the QCP moves to a higher value of $h$, as shown in Fig.~\ref{fig:model}(c). And comparing with the QCP of DQMC, $h_c=1.83(1)$, the QCP of EQMC is a slight different value, $h_c=1.84(1)$, since the position of the critical is a non-universal quantity sensitive to UV cutoffs. But the universal part of the quantum critical point, as will be discussed later, remain the same between DQMC and EQMC.

Furthermore, because fermions and Ising spins are coupled together, the Ising-spin clock phase immediately generates a spin-density-wave ordering in the fermionic sector with finite ordering wavevector $\mathbf{Q}$, which folds the Brillouin zone and renders a new FS with pockets as schematically shown in Fig.~\ref{fig:FS}(a). Near the QCP, as shown in our previous work~\cite{ZHLiu2017}, the quasi-particle at the tip of the FS pockets lose their coherence, forming the so-called hot spots~\cite{Metlitski2010a,Metlitski2010b,Schlief2017}.

As shown in Ref.~\onlinecite{ZHLiu2017}, with the help of cluster updates (Wolff, Swendsen-Wang and geometric-cluster)~\cite{Swendsen1987,Wolff1989,Heringa1998} and cumulative updates in the self-learning Monte Carlo method~\cite{liu2016self,liu2016fermion,Xu2016self,Nagai2017},
we were able to simulate systems as large as $L=30$ and temperature as low as $\beta=30$, and overcome the critical slowing down in the vicinity of QCP, to some extend. Up to such system sizes, non-Fermi-liquid around the hot spots can be clearly seen, moreover, the quantum critical scaling in the critical region can be investigated via dynamic susceptibility of the Ising spins, we found that the dynamic susceptibility scales as
\begin{align}
&\chi(T,h,\mathbf{q},\omega_n)= \nonumber\\
& \frac{1}{(c_{t}T+c_{t}' T^{2})+c_{h}|h-h_c|^{\gamma}+c_q |\mathbf{q}|^2 + (c_{\omega}\omega+c'_{\omega}\omega^{2})},
\label{eq:susceptibility}
\end{align}
where $c_t$, $c'_t$, $c_h$, $\gamma$, $c_q$, $c_{\omega}$ and $c'_{\omega}$ are determined from fitting the QMC data upto $L=\beta=30$. It is important to note that, at low temperature and frequency, the system encounters crossover behavior from $T^2$ to $T$ and from $\omega^2$ to $\omega$. This is consistent with the expectation of Hertz-Millis-Moriya theory~\cite{Hertz1976,Millis1993,Moriya1985}. However, we also note that to reveal the true IR physics of this itinerant antiferromagnetic QCP, $L=30$ and $\beta =30$ are clearly not large enough, as there are higher order perturbative RG calculations~\cite{Metzner2003,Abanov2004,Metlitski2010b,Schlief2017} suggesting the existence of anomalous dimenstion, i.e., the momentum-dependence of $\chi^{-1}$ is not $|\mathbf{q}|^2$ but $|\mathbf{q}|^{2-\eta}$, with finite anomalous dimension $\eta$. To address such question, we will have to go to even large system sizes and lower temperatures, and that is partially our motivation to develop the EQMC method in this work.

\subsection{Implementation of EQMC}
\label{sec:implement}

The general practice and step of EQMC have been outlined in Sections~\ref{sec:ideas} and ~\ref{sec:steps}. Here, we will discuss the model-specific implementation of the algorithm.

After the Fourier transformation, the kinetic energy in Eq.~\eqref{eq:HfM} becomes
\begin{equation}
  \label{eq:HfMk}
H_{f}=\sum_{\mathbf{k},\lambda,\sigma}[\epsilon(\mathbf{k})-\mu]c^{\dagger}_{\mathbf{k},\lambda,\sigma}c_{\mathbf{k},\lambda,\sigma},
\end{equation}
with $\epsilon(\mathbf{k})=-2t\cos(k_x)-4t\cos(\frac{\sqrt{3}}{2}k_y)\cos(\frac{1}{2}k_x)$. The bosonic part, $H_{b}$, will still be kept in the real space with size $N\times N$. The coupling term, $H_{fb}$, is then transformed as,
\begin{equation}
\label{eq:HfbMk}
H_{fb}=
-\xi\sum_{\mathbf{k},\mathbf{k'},\lambda}s^{z}(\mathbf{k}-\mathbf{k'})(c^{\dagger}_{\mathbf{k},\lambda,\uparrow}c_{\mathbf{k'},\lambda,\uparrow}-c^{\dagger}_{\mathbf{k},\lambda,\downarrow}c_{\mathbf{k'},\lambda,\downarrow}),
\end{equation}
where $s^{z}(\mathbf{q})=\frac{1}{N}\sum_{i}e^{-i\mathbf{q}\cdot\mathbf{r}_i}s^{z}_{i}$ plays the role of the the bosonic field $\phi_{k}$ in Eq.~\eqref{eq:phik}.

In the triangular lattice model, the ordering wave vectors $\pm\mathbf{Q}$ connect 6 pairs of hot spots (12 in total) on the FS.
We denote the 6 pairs of hot spots as $\{\mathbf{K}_{l}, \mathbf{K^\prime}_{l}\}$, with $l=1,2,\cdots,6$, as shown in Fig.~\ref{fig:FS} (a).
The distance between each two of the pair is one of the two ordering wave vectors, $\mathbf{K^\prime}_{l} - \mathbf{K}_{l} = \mathbf{Q}_l = \pm\mathbf{Q}$.
In the IR limit, only the fluctuations within a pair of hot spots $\{\mathbf{K}_{l}, \mathbf{K'}_{l}\}$ are important to the universal scaling behavior in the vicinity of QCP~\cite{Metzner2003,Abanov2003,Abanov2004,Loehneysen2007,Metlitski2010,Metlitski2010a,Metlitski2010b}.
Hence, to study this universal behavior, we draw one patch around each $\mathbf{K}_{l}$ and keep fermion modes therein, while neglect other parts of the BZ.
In this way, instead of the original $N=L \times L$ total points one needs to keep, we will only keep the $N_f=L_f\times L_f$ points inside each patch, as marked by the 12 leaf-like polygons in Fig.~\ref{fig:FS} (b) that covering the 12 hot spots.
Here, $L$ and $L_f$ denote the linear size of the original lattice and the size of the patch, respectively.

For the sake of bookkeeping, we label momenta in each patch using the relative difference $\mathbf{q}$ measured from the center of the patch: the real momentum is $\mathbf{k} = \mathbf{K}_l+\mathbf{q}$ or $\mathbf{k^\prime}=\mathbf{K^\prime}_l+\mathbf{q}$, respectively.
Correspondingly, we label the fermion mode in each pair of patches as $c_{\mathbf{q}l\lambda\alpha} = c_{\mathbf{K}_l+\mathbf{q},\lambda\alpha}$ and $c_{\mathbf{q}l\lambda\alpha}^\prime = c_{\mathbf{K}_l^\prime+\mathbf{q},\lambda\alpha}$, respectively.
In this way, $\mathbf{q}$ is a small momentum around zero.
Using this notation, the Hamiltonian in Eqs.~\eqref{eq:HfMk} and \eqref{eq:HfbMk} becomes
\begin{align}
H_{f} & = \sum_{\mathbf{q}l\lambda\sigma}\left\{
[\epsilon(\mathbf{q} + \mathbf{K}_l)-\mu]
c_{\mathbf{q}l\lambda\sigma}^{\dagger}c_{\mathbf{q}l\lambda\sigma}\right.\nonumber\\
&\quad\quad\quad\left.+[\epsilon(\mathbf{q}+\mathbf{K^\prime}_l)-\mu]
c_{\mathbf{q}l\lambda\sigma}^{\prime\dagger}c_{\mathbf{q}l\lambda\sigma}^\prime\right\}
 \label{eq:HfMq}\\
H_{fb} & = -\xi\sum_{\mathbf{q},\mathbf{q^\prime},l,\lambda}\left[
s^{z}(\mathbf{q}-\mathbf{q^\prime}+\mathbf{Q}_{l})
c_{\mathbf{q}l\lambda\alpha}^\dagger\sigma_{\alpha\beta}^zc_{\mathbf{q'}l\lambda\beta}^\prime
\right.\nonumber\\
&\quad\quad\quad+\left.
s^{z}(\mathbf{q}^\prime-\mathbf{q}-\mathbf{Q}_{l})
c_{\mathbf{q^\prime} l\lambda\alpha}^{\prime\dagger}
\sigma_{\alpha\beta}^zc_{\mathbf{q}l\lambda\beta}\right].
\label{eq:HfbMq}
\end{align}
where the sum over $\mathbf{q}$ are in the patches around the hot spots.

We notice that, comparing to Eq.~\eqref{eq:HfbMk}, only scattering between two modes in a pair of patches is kept in Eq.~\eqref{eq:HfbMq}. This is because the distance between two momentum points in a pair of patches can be expressed as
\begin{align}
\mathbf{k}-\mathbf{k'}&=\mathbf{q}-\mathbf{q'}+(\mathbf{K}_{l}-\mathbf{K'}_{l}) \nonumber\\
&=\mathbf{q}-\mathbf{q'}+\mathbf{Q}_{l},
\end{align}
which is approximately equal to $\mathbf{Q}_l=\pm\mathbf{Q}$.
Hence, the momentum transfers in these scattering correspond to low-energy spin fluctuations, and such scattering processes are relevant to the IR universal behavior.
On the contrary, other scattering processes have momentum transfers different from $\pm \mathbf{Q}$, and thus are not relevant to the IR universal behavior.
This approximation is discussed in the general setting of EQMC in Sec.~\ref{sec:steps}.

We now outline the construction of the $\mathbf{K}$ and $\mathbf{V}$ matrices (defined in Sec.~\ref{sec:steps}) using Eqs.~\eqref{eq:HfMq} and \eqref{eq:HfbMq}, which are needed in computing the fermion determinant.
In total, we keep $12\times L_f\times L_f$ fermion modes.
As discussed in Sec.~\ref{sec:steps}, the matrices are block-diagonalized into 6 blocks, corresponding to the six pairs of hot spots.

First, the fermion dispersion in Eq.~\eqref{eq:HfMq} is converted into the following $\mathbf{K}$ matrix,
\begin{equation}
\mathbf K_{\sigma}\,=\,\left[\begin{array}{cccc}
t_{1\sigma}\\
 & t_{2\sigma}\\
 &  & \ddots\\
 &  &  & t_{l\sigma}
\end{array}\right]
\label{eq:Tmatrix}
\end{equation}
with
\begin{equation}
t_{l\sigma}\,=\,\begin{bmatrix}\epsilon(\mathbf{q}+\mathbf{K}_{l})-\mu & 0 & 0 & 0\\
0 & \ddots & 0 & 0\\
0 & 0 & \epsilon(\mathbf{q'}+\mathbf{K'}_l)-\mu & 0\\
0 & 0 & 0 & \ddots
\end{bmatrix}.
\label{eq:tlmatrix}
\end{equation}
Here, $\mathbf{K}$ is divided into 6 blocks, each corresponds to one pair of patches.
Each block is further divided into two halves, representing patches around $\mathbf{K}_l$ and $\mathbf{K'}_l$, respectively.
Within each half, each entry corresponds to one momentum point $\mathbf{q}$ in the patch.


Second, the fermion-spin coupling in Eq.~\eqref{eq:HfbMq} is converted into the following $\mathbf V$ matrix,
\begin{equation}
V_{\sigma}\,=\,\left[\begin{array}{cccc}
v_{1\sigma}\\
 & v_{2\sigma}\\
 &  & \ddots\\
 &  &  & v_{l\sigma}
\end{array}\right],
\label{eq:Vmatrix}
\end{equation}
where $v_{l}$ is an off-diagonal block matrix. The matrix element in $v_{l}$ represent the interaction within one hot-spot pair $\{\mathbf{K}_{l}, \mathbf{K'}_{l}\}$, whereas the other matrix element in $V$ matrix, representing the momentum modes interaction between diffrerent hot pots pair $\{\mathbf{K}_{l}, \mathbf{K'}_{l'}\}$, with $l \neq l'$, are set to zero under our IR limit approximation.

The structure of each $v_{l}$ is
\begin{equation}
v_{l\sigma}\,=\,\begin{bmatrix}0 & 0 & s(\mathbf{q}-\mathbf{q}'+\mathbf{Q}_{l}) & \cdots\\
0 & 0 & \vdots & \ddots\\
s(\mathbf{q}'-\mathbf{q}-\mathbf{Q}_{l}) & \cdots & 0 & 0\\
\vdots & \ddots & 0 & 0
\end{bmatrix}
\label{eq:vlmatrix}
\end{equation}
with matrix elements in the two diagonal blocks are zero since there is no scattering within the leaf-like polygon of one hot spot $\mathbf{K}_{l}$ or $\mathbf{K'}_{l}$, however, the matrix element in the two off-diagonal blocks are non-zero since it is the scatterings between $\mathbf{K}_{l}$ and $\mathbf{K'}_{l}$ , i.e., $s(\mathbf{q}-\mathbf{q'}+\mathbf{Q}_{l})$ in Eq.~\eqref{eq:vlmatrix}. Since $s_i^z$ are real, $s(\mathbf{q}-\mathbf{q}'+\mathbf{Q}_{l})=s^{*}(\mathbf{q}'-\mathbf{q}-\mathbf{Q}_{l})$ and $v_{l}=v_{l}^{\dagger}$, such that one only needs to handle the matrix elements coming from $s(\mathbf{q}-\mathbf{q}'+\mathbf{Q}_{l})$.

Once having both block diagonal matrices of $\mathbf{K}_{\sigma}$ and $\mathbf{V}_{\sigma}$ in the fermion determinant in the configurational weight in Eq.~\eqref{eq:Wphi}, the Fermion Green's function matrice also becomes block diagonal,
\begin{equation}
G_{\sigma}\,=\,\left[\begin{array}{cccc}
G_{1\sigma}\\
 & G_{2\sigma}\\
 &  & \ddots\\
 &  &  & G_{l\sigma}
\end{array}\right],
\end{equation}
this means we can update different pair of hot spots, namely, different $G_{l\sigma}$, independently.
As claimed in Sec.~\ref{sec:complexity}, this block structure not only reduces memory usages, but also reduces computational cost and makes it easier to parallelize the simulation.


Finally, we describe the update scheme used in our simulation.
As outlined in Sec.~\ref{sec:ideas}, the local-update algorithm, which is widely used in conventional DQMC simulations~\cite{BSS1981,Hirsch_1983,Hirsch_1985,White1989,AssaadEvertz2008}, becomes inefficient in EQMC simulations.
This is a result of the fact that the form of Eq.~\eqref{eq:HfbMq} is nonlocal in the momentum space.
Consequently, flipping one spin $s_i^z$ in a spin configuration changes all rows and columns in the $\mathbf V$ matrix, and all the Fourier components $s(\bm q)$ are affected.
As a result, the fast-update method, which computes the ratio between two determinants differing only one row and one column with complexity $O(N)$ (where $N$ is the size of the matrix), no longer applies.

Instead, we use global-update algorithms.
In general, these algorithms stochastically propose a completely new spin configuration with a relatively low cost.
The new configuration is then either accepted or rejected, following the detailed balance principle.
Assuming that starting from a configuration $\{s_i^z\}$, the probability of constructing the configuration $\{\tilde s_i^z\}$ is $S\left(s^z\rightarrow\tilde s^z\right)$, then the detailed balance principle requires that probability of accepting this new configuration is
\begin{equation}
\label{eq:dbp}
\alpha\left(s^z\rightarrow\tilde s^z\right) =
\min\left\{\frac{S\left(\tilde s^z\rightarrow s^z\right)}
{S\left(s^z\rightarrow\tilde s^z\right)}
\frac{W[\tilde s^z]}{W[s^z]}, 1\right\}.
\end{equation}
A good global update satisfies two conditions:
(i) it generates a new configuration sufficiently different from the original one, such that it has almost no correlation with the previous one.
(ii) the acceptence ratio computed from Eq.~\eqref{eq:dbp} is close to one.
These two conditions guarentee that the autocorrelation time in the generated Markov chain is on the order of one.

In our simulations, we use two types of global updates:
First, we use cluster updates guided by the bosonic part of the Hamiltonian $H_b$ only.
In this way, a new configuration is constructed using well-known algorithms, including Wolff~\cite{Wolff1989}, Swendsen-Wang~\cite{Swendsen1987} and geometric cluster update schemes~\cite{Heringa1998}.
These cluster-update algorithms imply that the probability $S\left(s^z\rightarrow\tilde s^z\right)$ is determined by $H_b$,
\begin{equation}
\label{eq:SWb}
\frac{S\left(s^z\rightarrow\tilde s^z\right)}
{S\left(\tilde s^z\rightarrow s^z\right)}
=\frac{W_b[\tilde s^z]}{W_b[s^z]}.
\end{equation}
The acception probability can then be computed using Eq.~\eqref{eq:dbp}.

Second, we make use of the cumulative update in the self-learning Monte Carlo method~\cite{liu2016self,liu2016fermion,Xu2016self}. To use this method, one first learns an effective bosonic model $H_{\text{eff}}$ of the total system from the trainning steps with the configurational weight generated with other update methods. The effective model is in general nonlocal both in space and time, but since it is a bosonic model, we can update the entire space-time configuration through it locally with relatively low computational cost (since this step is done without evoking the calculation of Fermi determinant).
This step is done repeatedly, until a completely new bosonic configuration with little correlation with the previous one is generated.
The collection of all these local steps is called the ``cumulative update'', and serves as the construction process of a global update.
Similar to Eq.~\eqref{eq:SWb}, the selection probability is given by the learned effective model,
\begin{equation}
\label{eq:SWeff}
\frac{S\left(s^z\rightarrow\tilde s^z\right)}
{S\left(\tilde s^z\rightarrow s^z\right)}
=\frac{W_{\text{eff}}[\tilde s^z]}{W_{\text{eff}}[s^z]}.
\end{equation}

Both types of global updates construct a new configuration with little correlation with the previous one. Furthermore, plugging Eqs.~\eqref{eq:SWb} and \eqref{eq:SWeff} into Eq.~\eqref{eq:dbp}, we can see that the resulting acceptance ratio $\alpha$ is close to one, if the bosonic models guiding the construction with weight $W_b$ and $W_{\text{eff}}$, respectively, give good approximations to the full weight $W$.
Hence, these update algorithms satisfy the conditions listed above, and thus can generate statistically independent samples with only a few ($O(1)$) update steps.

In these global-update algorithms, the step dominating the computational cost is computing the true weight $W[s^z]$, in order to determine $\alpha$ in Eq.~\eqref{eq:dbp}.
As discussed in Sec.~\ref{sec:complexity}, this in turn is dominated by computing the fermion determinant, which costs $O(\beta N_f^3)$.
It is easy to check that constructing a new configuration, using either $W_b$ or $W_{\text{eff}}$, is much faster in comparison. From the above analysis, we can conclude that the computational complexity for EQMC is $O( \beta N_f^3)$, at least $(N/N_f)^3$ times faster than that of the typical DQMC's $O( \beta N^3)$. And because we can easily have $N/N_f \sim 10$ (in the results in Sec.~\ref{sec:results}, we have $\frac{N}{N_f}=36$), thousands times of speedup of EQMC over DQMC can be readily accessed, as demonstrated in the next section.

\begin{figure}[htp]
\centering
\includegraphics[width=\columnwidth]{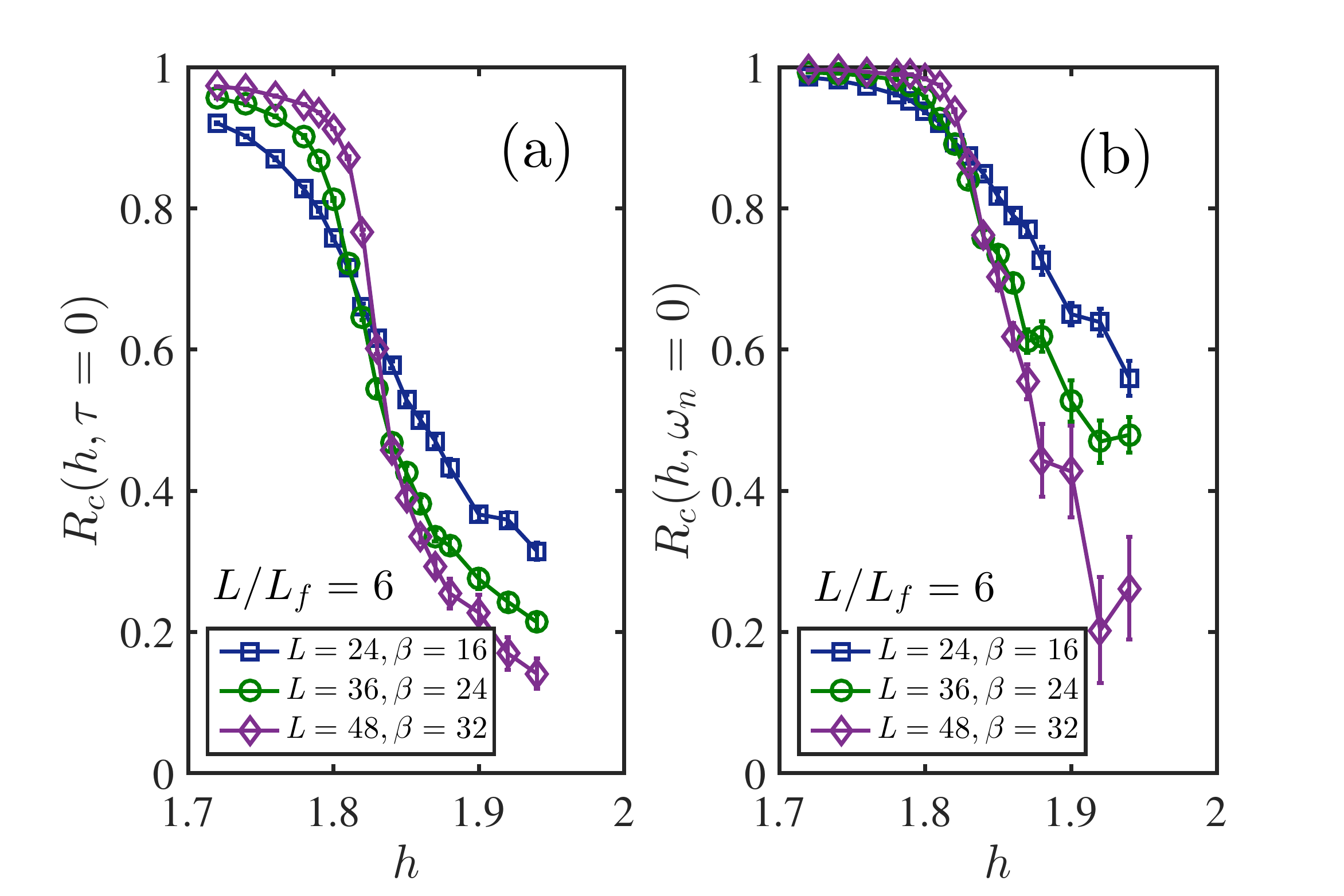}
\caption{(a) Correlation ratio $R_c(h,\tau=0)$ (b) $R_c(h,\omega=0)$ of the Ising magnetic order in clock phase, as a function of $h$, with $L_f=4,6,8$ (corresponding to $L=24,36,48$) and $\beta=4\times L_f$, obtaind from EQMC simulation. As $L$ increases, the crossing point is found to be converged at $h_c=1.84(1)$.}
\label{fig:correlationratio}
\end{figure}

\subsection{Results}
\label{sec:results}
To demonstrate the power and efficiency of EQMC, we investigate the triangle lattice model described above, and in particular, pay attention to the scaling behavior in the vicinity of the antiferromagnetic quantum critical point. We chose to fix the ratio of $L/L_f=6$ such that $N/N_f=36$, and the size of the leaf-patches $L_f=4,6,8$, and fixed $\beta=4\times L_f$. This means that we can simulate the original model in Fig.~\ref{fig:model} (a) upto size $48\times48\times32$ ($L\times L\times \beta$), almost 3 times of the largest size ever simulated in the DQMC~\cite{ZHLiu2017}.

In our previous work Ref.~\onlinecite{ZHLiu2017}, the bosonic susceptibilities $\chi(T, h, \bm q, \omega)$ close to the QCP, revealed with $L=\beta=30$, is fitted to the form of Eq.~\eqref{eq:susceptibility}.
In particular, at low $\omega$, $\chi^{-1}(0, 0, 0, \omega)$ exhibits crossover behavior from $\omega^2$ to $\omega$. It also scales with $\bm q$ as $\chi^{-1}(0, 0, \bm q, 0)\propto|\bm q|^2$.

\begin{figure*}[htp]
\centering
\includegraphics[width=\textwidth]{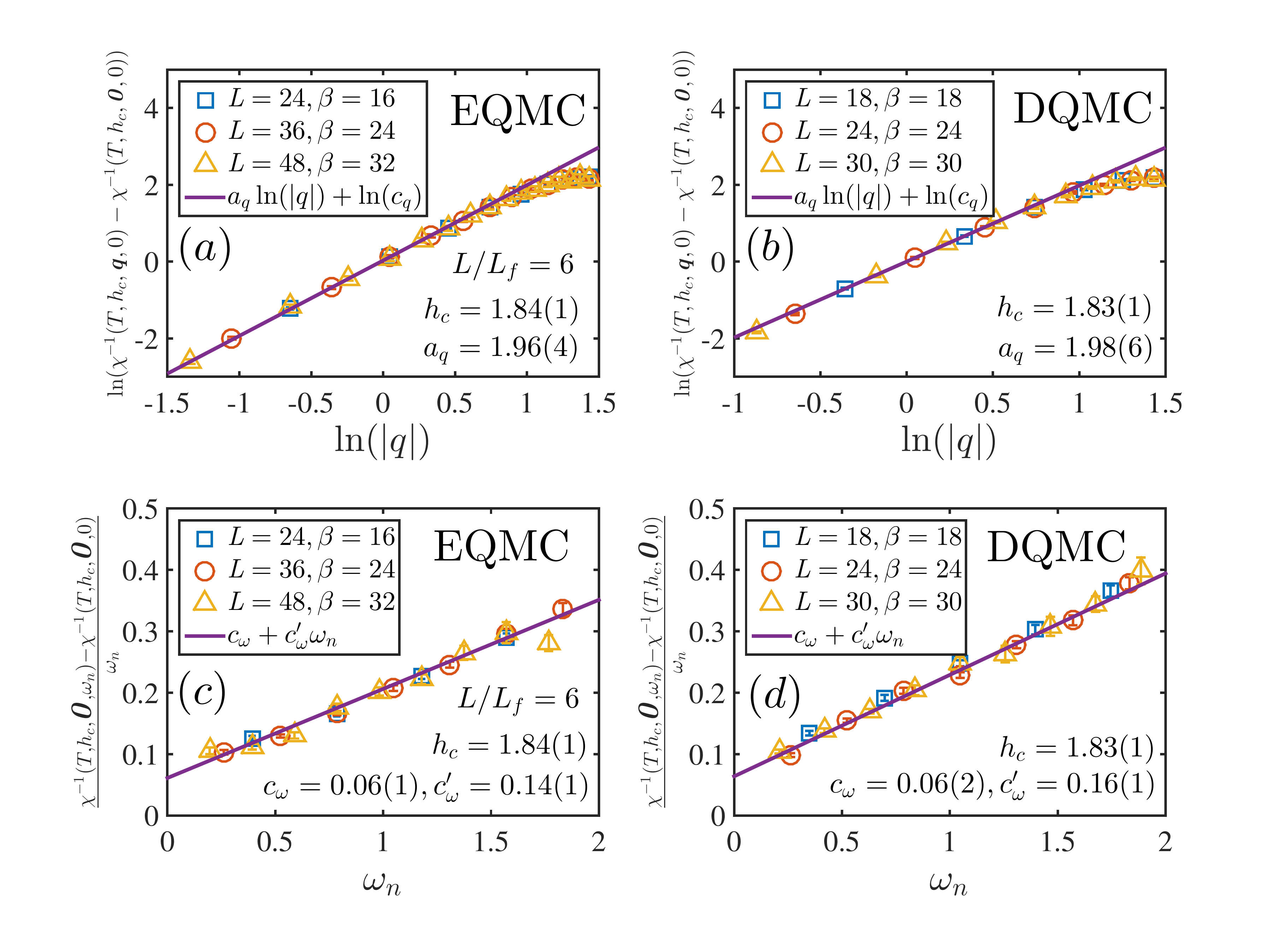}
\caption{$|\mathbf{q}|$, $\omega$ and $T$ dependence of the bosonic susceptibilities $\chi(T,h,\mathbf{q},\omega)$ at the itinerant QCP $h=h_c$. Comparison of the EQMC results with $L$ upto 48 and $\beta$ upto 32 (a) and (c), with the previous DQMC results with $L$ and $\beta$ upto 30 (b) and (d). The scaling behaviors with system sizes as large as $48\times48\times32$ ($L\times L\times \beta$) in EQMC are fully consistent with the form in Eq.~\ref{eq:susceptibility}. The universal quantum critical scaling has been successfully captured by EQMC.}
\label{fig:chi-qw-analysis}
\end{figure*}

As we discuss in the Sec.~\ref{sec:ideas}, the model we simulated with EQMC [Eqs.~\eqref{eq:HbM}, \eqref{eq:HfMq} and \eqref{eq:HfbMq}], which includes only the k-points inside the hot spot patches, is different from the orignial model in their nonuniversal properties, due to approximation made in EQMC. This is shown in Fig.~\ref{fig:model} (c), where the location of the QCP can shift from $h_c=1.83(1)$ obtained from DQMC in Ref.~\onlinecite{ZHLiu2017} to $h_c=1.84(1)$ obtianed from EQMC. To determine the position of the new $h_c$, we measure the correlation ratio~\cite{Pujari2016,Xu2017} of the Ising spins as a function of $h$. We first calculate the magnetic susceptibility in the EQMC simulation
\begin{equation}
\chi(T,h,\mathbf{q},\omega_n)=\frac{1}{L^2}\sum_{ij}\int^{\beta}_{0}d\tau \exp^{i(\omega_n\tau -\mathbf{q}\cdot\mathbf{r}_{ij})}\langle s^{z}_{i}(\tau)s^{z}_{j}(0)\rangle,
\end{equation}
then construct the correlation ratio both from the equare-time susceptibility (magnetic structure factor)
\begin{equation}
\label{eq:Rctau0}
R_c(h,\tau=0)=1-\frac{\chi(T,h,\mathbf{Q}+d\vec{k},\tau=0)}{\chi(T,h,\mathbf{Q},\tau=0)},
\end{equation}
and from the zero frequency susceptibility
\begin{equation}
\label{eq:Rcomega0}
R_c(h,\omega=0)=1-\frac{\chi(T,h,\mathbf{Q}+d\vec{k},\omega_n=0)}{\chi(T,h,\mathbf{Q},\omega_n=0)},
\end{equation}
where $d\vec{k}$ can be chosen as $\vec{b}_1=\frac{2\pi}{L}(1,\frac{1}{\sqrt{3}})$ or $\vec{b}_2=\frac{2\pi}{L}(0,2\sqrt{3})$, which corresponded to the minimum distance in momentum space.

The results are shown in Fig.~\ref{fig:correlationratio} (a) and (b). It is clear that with $L_f=4,6,8$, which means $L=24,36,48$, the crossing point of the correlation ratio is converged to $h_c = 1.84(1)$ from EQMC simulations.

What is more important, is that the universal properties of the QCP, shall remain intact with EQMC. Therefore, we further explore this assessment by means of analyzing the various divergences of $\chi(T,h,\mathbf{q},\omega_n)$ in the form of Eq.~\eqref{eq:susceptibility} in the qunatum critical region. The results are shown in Fig.~\ref{fig:chi-qw-analysis}.

We first look at the $\mathbf{q}$ dependence, as shown in Fig.~\ref{fig:chi-qw-analysis} (a), the momentum $|\mathbf{q}|$ is measured with respect to the hot spot position $\mathbf{K}$. We plot the susceptibility data by substraction the finite temperature background as, $\chi^{-1}(T,h_c,|\mathbf{q}|,0)-\chi^{-1}(T,h_c,0,0)=c_q |\mathbf{q}|^{a_q}$, and fit the curve to obtain the coefficient $c_q$ and the power $a_q$, as shown in the red solid line. With the system size as large as $L=48$, the $\chi^{-1}\sim |\mathbf{q}|^2$ behavior, clearly manifests, with $a_q=1.96(4)$. These results are consistent with that obtained in our previous work~\cite{ZHLiu2017}, where the anomalous dimension $\eta$ of this critical point, in the form of $\chi^{-1} \sim |\mathbf{q}|^{2-\eta}$, is zero within errorbar. For the sake of completeness, the DQMC results from Ref.~\cite{ZHLiu2017} are also shown in Fig.~\ref{fig:chi-qw-analysis} (b).

As for the frequency dependence in $\chi$, it is also consistent with our previous conclusion in Ref.~\onlinecite{ZHLiu2017}. As shown in Fig.~\ref{fig:chi-qw-analysis} (c), we analyze the $(\chi^{-1}(T,h_c,0,\omega)-\chi^{-1}(T,h_c,0,0))/\omega$, to substract the finite temperature background. From the plot, it is clear that there is a finite intercept at $\omega=0$, and  from the fit, the form of $\chi^{-1} \sim c_\omega \omega + c'_\omega \omega^2$ can be observed, with $c_\omega=0.06(1)$ and $c'_\omega=0.14(1)$, fully consistent with our previously determined the coefficients with DQMC, as also shown in Fig.~\ref{fig:chi-qw-analysis} (d).

The results in Fig.~\ref{fig:chi-qw-analysis} conclude that the form in Eq.~\eqref{eq:susceptibility} accurately describe this itinerant antiferromagnetic QCP. This form is different from the bare $(2+1)$D O(2) universality which is the description of the bare bosonic problem~\cite{Isakov2003,YCWang2017}, and is consistent with the Hertz-Millis-Moriya description~\cite{Hertz1976,Millis1993,Moriya1985}. However, the proposed anomalous dimension of this antiferromagnetic itinerant QCP with higher-order perturbative RG calculations~\cite{Abanov2004}, has not been observed with system size as large as $L=48$.

\section{Discussions}
\label{sec:conclusion}
The elective momentum ultra-size quantum Monte Carlo method (EMUS-QMC or EQMC), developed in this work, pave the way of performing quantum Monte Carlo simulations for larger system sizes and lower temperature, such that the genuine IR physics of many interesting yet difficult strongly correlated electron problems, exemplified with the itinerant antiferromagnetic quantum critical point here, are now ready to be explored. EQMC manage to reduce the notorious $O(\beta N^3)$ computational complexity of the conventional DQMC down to $O(\beta N^3_f)$, where $N_f$ is the important momentum points inside the patch around hot spots. Since $\frac{N}{N_f} \sim 10$ is easy to design, as shown in this work (here we have $\frac{N}{N_f}=36$), a speedup to the order of $10^3$ is achieved. In the benchmark example we demonstrated, the system sizes of $48\times48\times32$ ($L\times L\times \beta$) can be comfortably simulated without the great computational effort spent in previous work that only $30\times30\times30$ can be accessed.

Moreover, the speedup and model flexibility offered by EQMC opens up opportunities to study other interesting universality classes of fermion QCPs, in particular, those that are hard to be realized in a lattice model because the UV completion of the desired IR fermion mode would require many fermion modes in the BZ.
For example, a similar antiferromagnetic square-lattice spin-fermion model can be used to study an itinerant QCP with a $\mathbb Z_2$ symmetry, where the EQMC method offers a speedup similar to what is demonstrated in this work, and existence or absence of the anomalous dimension in that case can be verified~\cite{ZiHongLiu201808}.
EQMC can also be applied to models where Dirac fermions interact with bosonic modes~\cite{Xu2016a,He2017}. Comparing to DQMC simulations using a honeycomb-lattice model or a $\pi$-flux square-lattice model, EQMC offers a speedup as only fermion modes with the linear dispersion of Dirac fermions are included~\cite{Lang2018}, and therefore much larger system sizes and lower temperatures can be accessed, to resolve the present difference between numerical simulation on finite lattice and analytical field-theory calculations at the theromdynamic limit on the critical exponents of these models~\cite{Iliesiu2016,Iliesiu2017,Zerf2017,Ihrig2018}. Furthermore, EQMC can be applied to investigate interaction effects on surface states of three-dimensional (3D) topological insulators and topological superconductors, which cannot be realized in a two-dimensional (2D) lattice model. With EQMC, we can simulate the 2D surface modes without the corresponding 3D bulk, thus greatly reduce the computational cost.


\section{Acknowledgement}
We acknowledge valuable discussions with Anders Sandvik, Andrey Chubukov, Sung-Sik Lee, Michael M. Scherer and Fakher F. Assaad on various subjects of itinerant quantum critcality. ZYM is in debt to Xi Dai and Tao Xiang for illuminating discussions in which the idea of EQMC is further crystalized. ZHL, XYX and ZYM acknowledge support from the Ministry of Science and Technology of China through the National Key Research and Development Program (Grant No. 2016YFA0300502), the Strategic Priority Research Program of the Chinese
Academy of Sciences (Grant No. XDB28000000), and the National Science Foundation of China (Grant No. 11421092,11574359 and 11674370). YQ acknowledges support from Minstry of Science and Technology of China under grant numbers 2015CB921700, and from National Science Foundation of China under grant number 11874115. We thank the Center for Quantum Simulation Sciences in the Institute of Physics, Chinese Academy of Sciences and the Tianhe-1A platform at the National Supercomputer Center in Tianjin for their technical support and generous allocation of CPU time. XYX also acknowledges the support of HKRGC through grant C6026-16W. K.S. acknowledges support from the National Science Foundation under Grant No. PHY1402971 and the Alfred P. Sloan Foundation.

\bibliography{FQMC}

\end{document}